\documentclass{ps}\usepackage[]{graphicx}\usepackage[]{color}
\makeatletter
\def\maxwidth{ %
  \ifdim\Gin@nat@width>\linewidth
    \linewidth
  \else
    \Gin@nat@width
  \fi
}
\makeatother

\definecolor{fgcolor}{rgb}{0.345, 0.345, 0.345}

\usepackage{framed}
\makeatletter
 {\par\unskip\endMakeFramed%
 \at@end@of@kframe}
\makeatother

\definecolor{shadecolor}{rgb}{.97, .97, .97}
\definecolor{messagecolor}{rgb}{0, 0, 0}
\definecolor{warningcolor}{rgb}{1, 0, 1}
\definecolor{errorcolor}{rgb}{1, 0, 0}
\newenvironment{knitrout}{}{} % an empty environment to be redefined in TeX

\usepackage{alltt}
\usepackage[utf8]{inputenc}
\usepackage[english]{babel}
\usepackage[T1]{fontenc}
\usepackage{amsmath}
\usepackage{amsfonts}
\usepackage{amssymb}
\usepackage{amsthm}
\usepackage{import}
\newcommand{\E}[1]{\operatorname{E}\left[ #1 \right]}
\newcommand{\var}[1]{\operatorname{var}\left[ #1 \right]}

\newcommand{\cv}[1]{\operatorname{CV}\left[ #1 \right]}
\newcommand{\proba}[1]{\operatorname{P}\left[ #1 \right]}
\newcommand{\card}{\operatorname{card}}
\usepackage{bbm}
\newcommand{\one}{\mathbbm{1}}
\newcommand{\X}{\ensuremath{\mathbf{X}}}

\renewcommand{\u}{\mathbf{u}}
\newcommand{\U}{\ensuremath{\mathbf{U}}}

\newcommand{\R}{\ensuremath{\mathbb{R}}}
\newcommand{\N}{\ensuremath{\mathbb{N}}}
\newcommand{\p}{\ensuremath{\widehat{p}}}
\newcommand{\pmc}{\ensuremath{\widehat{p}_\text{MC}}}

\usepackage{xparse}
\NewDocumentCommand{\Sum}{O{i = 1}O{N}}{\sum \limits_{#1}^{#2}}
\NewDocumentCommand{\dSum}{O{i = 1}O{N}}{ \dfrac{1}{#2} \sum \limits_{#1}^{#2}}
\renewcommand{\l}{\left(}
\renewcommand{\r}{\right)}

\usepackage{letltxmacro}
\LetLtxMacro{\oldsubsection}{\subsection}
\NewDocumentCommand\labsubsec{ s m }{% s = star, m = mandatory arg
   \IfBooleanTF{#1}{%
      \oldsubsection*{#2}%
   }{%
      \oldsubsection{#2}%
   }%
    \label{ss:#2}
}
\renewcommand{\subsection}{\labsubsec}
\LetLtxMacro{\oldsection}{\section}
\NewDocumentCommand\labsec{ s m }{% s = star, m = mandatory arg
   \IfBooleanTF{#1}{%
      \oldsection*{#2}%
   }{%
      \oldsection{#2}%
   }%
    \label{s:#2}
}
\renewcommand{\section}{\labsec}
\LetLtxMacro{\oldbiblio}{\bibliography}
\renewcommand{\bibliography}[1]{%
	\renewcommand{\section}{\oldsection}%
	\oldbiblio{#1}%
}
\let\originaleqref\eqref
\renewcommand{\eqref}{Eq.~\originaleqref}
\newcommand{\cdf}{\textit{cdf }}
\newcommand{\iid}{\textit{iid. }}
\newcommand{\appenproof}{\oldsection*{Appendix}}
\newcommand{\eg}{\textit{e.g. }}

\usepackage{graphicx}
\usepackage{subfig}
\usepackage[left=2cm,right=2cm,top=2cm,bottom=2cm]{geometry}
\usepackage{algpseudocode}
\usepackage{algorithm}
\usepackage[inline]{enumitem}
\usepackage[numbers]{natbib}
\usepackage{color}
\usepackage{stmaryrd}
\usepackage{answers}
\usepackage{tikz}
\Newassociation{prooftheo}{demtheo}{ann}

\Newassociation{proofprop}{demprop}{ann}

\Newassociation{prooflem}{demlem}{ann}

\Newassociation{proofcoro}{demcoro}{ann}

\Opensolutionfile{ann}[demo]
\begin{document}
\author{Clément Walter}
\address{CEA, DAM, DIF, F-91297 Arpajon, France}
\secondaddress{Laboratoire de Probabilit\'es et Mod\`eles Al\'eatoires, Universit\'e Paris Diderot, 75205 Paris Cedex 05, France}
\thanks{This work was partially supported by ANR project Chorus}
\title{Rare Event Simulation and Splitting for Discontinuous Random Variables}

\begin{abstract}
Multilevel Splitting methods, also called Sequential Monte-Carlo or \emph{Subset Simulation}, are widely used methods for estimating extreme probabilities of the form $\proba{ S(\U) > q }$ where $S$ is a deterministic real-valued function and $\U$ can be a random finite- or infinite-dimensional vector. Very often, $X := S(\U)$ is supposed to be a continuous random variable and a lot of theoretical results on the statistical behaviour of the estimator are now derived with this hypothesis. However, as soon as some threshold effect appears in $S$ and/or $\U$ is discrete or mixed discrete/continuous this assumption does not hold any more and the estimator is not consistent.

In this paper, we study the impact of discontinuities in the \cdf of $X$ and present three unbiased \emph{corrected} estimators to handle them. These estimators do not require to know in advance if $X$ is actually discontinuous or not and become all equal if $X$ is continuous. Especially, one of them has the same statistical properties in any case. Efficiency is shown on a 2-D diffusive process as well as on the \emph{Boolean SATisfiability problem} (SAT).
\end{abstract}

\subjclass{65C05, 65C60, 62L12, 62N02}

\keywords{Rare Event Simulation, Multilevel Splitting, RESTART, Sequential Monte Carlo, Extreme Event Estimation, Counting, Last Particle Algorithm}

\maketitle

\section*{Introduction}
In the context of reliability analysis, one is often concerned with the estimation of extreme quantile or probability. Indeed one goal of uncertainty quantification is to estimate the probability of failure of a given system and inversely, quantile estimation helps defining guidelines to insure a \emph{good} behaviour of the system with a given probability of failure. Usually, the system is considered as a \emph{blackbox} (often a complex numerical code) which returns a real value defining its \emph{state}. According to this output, it is then considered as working properly or not.

Formally, the problem can be written as follows: let $\U$ be a random finite- or infinite-dimensional vector with known distribution $\mu^U$ and $S$ a \emph{performance function} (the computer code for instance), one seeks for estimating $p$ given $q$ (or $q$ given $p$) such that $p = \proba{S(\U) > q}$. The main difficulties arise from the fact that 
\begin{enumerate*}[label=\arabic*)]
\item the sought probability or quantile is extreme, say $p < \ensuremath{10^{-5}}$ and
\item the computer code is very time consuming.
\end{enumerate*}
All together, these two characteristics prevent from using a naive Monte-Carlo estimator:
\begin{equation} \label{eq:def monte carlo estim}
\pmc = \dfrac{1}{N} \Sum[i = 1][N] S(\U_i)
\end{equation}
with $(\U_i)_i$ an \emph{iid} sample with distribution $\mu^U$ because $\cv{\pmc}^2 \approx (Np)^{-1}$, which means that one would require $N = 10^2/p$ to get a coefficient of variation of $10\%$.

In this scope other statistics have been defined to get low variance estimators. Among them the Multilevel Splitting method (MS) \citep{kahn1951estimation,glasserman1999multilevel,rosenbluth2004monte,garvels2000splitting} rewrites the sought probability using the Bayes' rule and a finite sequence of increasing thresholds $(q_i)_{i=0}^m$ such that $q_0 = - \infty$ and $q_m = q$:
\begin{equation}\label{eq:definiton mutlilevel splitting}
\proba{S(\U) > q} = \proba{S(\U) > q_m \mid S(\U) > q_{m-1}} \times \cdots \times \proba{S(\U) > q_2 \mid S(\U) > q_1} \proba{S(\U) > q_1}
\end{equation}
From \eqref{eq:definiton mutlilevel splitting} the goal is then to estimate \emph{separately} each conditional probability with a crude Monte Carlo. Hence, the sequence $(q_j)_j$ has to be chosen such that each probability is \emph{not too small} to make its estimation with crude Monte-Carlo \emph{feasible} \citep{au2001estimation,rubinstein2010randomized}. The variance of the estimator depends on the choice of this sequence and especially it is known that the conditional probabilities should be all equal to minimize it \citep{cerou2012sequential}. A typical MS algorithm works as follows:
\begin{enumerate}
\item Sample a Monte-Carlo population $(\U_i)_i$ of size $N$; $j = 0$
\item Estimate the conditional probability $\proba{S(\U) > q_{j+1} \mid S(\U) > q_j}$ \label{line:MS scheme}
\item Resample the $\U_i$ such that $S(\U_i) \leq q_{j+1}$ conditionally to be greater than $q_{j+1}$ (the other ones do not change)
\item $j \gets j + 1$ and repeat until $j = m$
\end{enumerate}

When it is not possible to infer such a sequence $(q_j)_j$, it is usual to define it \emph{on-the-fly} while the algorithm is running and this is known as Adaptive Multilevel Splitting (AMS). The sequence is built either by fixing the conditional probabilities to be all equal to some given value $p_0 \in (0.1, 0.5)$ \citep{au2001estimation}, or by using the $k^{th}$ order statistics and so to estimate the corresponding probability by $1 - k/N$. In the first case, $(q_j)_j$ is then a sequence of estimated quantiles with crude Monte Carlo while in the second case it is a sequence of arbitrarily chosen thresholds and the corresponding probabilities are estimated with crude Monte Carlo. Hence the first option leads to a bias in the final estimator \citep{cerou2007adaptive,cerou2012sequential} while the latter produces an unbiased estimator for any $k$ \citep{brehier2014analysis}. The link with Interacting Particle Systems \citep{del2006sequential} allowed for deriving a lot of theoretical results on the optimal number of subsets, the statistical behaviour of the estimator and the impact of the adaptive method \citep{cerou2014fluctuation,beskos2013convergence}. Furthermore, the special case of the Last Particle Algorithm (AMS with $k=1$) has gained a lot of attention recently. It has the smallest variance amongst all AMS \citep{brehier2014analysis}; especially \citet{simonnet2014combinatorial} and \citet{guyader2011simulation} showed that the random number of iterations of the algorithm follows a Poisson law when $X$ is continuous. Moreover \citet{walter2014point}, following \citet{huber2011using}, brought an original insight in terms of a random walk of the real-valued random variable $X = S(\U)$ linked with a Poisson Process with parameter 1. Indeed the estimator is found to be the Minimal Variance Unbiased Estimator (MVUE) of the exponential of a parameter of a Poisson law with $N$ \iid realisations of such random variables. As a consequence it turns it into the optimal (minimal variance) parallel Multilevel Splitting estimator \citep{walter2015moving, walter2015rare}.

However, all these results assume that the \cdf of $X$ is continuous and little is known about the impact of using such strategies if this assumption does not hold. This case can happen if for example there is some threshold effect in $S$ and/or if $\U$ is discrete or mixed discrete/continuous \citep{cerou2011use,rubinstein2009gibbs,rubinstein2010randomized,rubinstein2012splitting}. Recently, \citet{simonnet2014combinatorial} showed that in the case of the Last Particle Algorithm the random number of iterations is indeed a mixture of independent Poisson and negative binomial laws while in the continuous case, it is only a Poisson law. Unfortunately he could not derive a general unbiased estimator from this result \citep[Theorems 4 and 5]{simonnet2014combinatorial}. The main problem comes from the fact that the equality $\forall x \in \R, \; \proba{X > x} = \proba{X \geq x}$ does not hold any more. \citet{rubinstein2009gibbs} already noticed that one should pay attention to the fact that the root $q_{i}$ of $\proba{S(\U) \geq q_{i} \mid S(\U) \geq q_{i-1}} = p_0$ may not be unique \citep[Remark 6.1]{rubinstein2009gibbs}, \citep[Remark 2.6]{botev2008efficient}. He then derived some guidelines for an appropriate adaptive choice of the $(q_i)_i$ in this case but concludes that the algorithm can eventually fail to estimate the sought probability (it stops at an intermediate level \citep[see][Remark 6.3]{rubinstein2009gibbs} or returns $0$ \citep{amrein2011variant}). Finally, \citet{cerou2011use} suggested to use for the SAT problem an auxiliary continuous random variable $Y$ such that $\proba{Y > q} = \proba{S(\U) > q}$ and showed practical improvement. This strategy is also used by \citet{huber2011using} for the Ising model but there are always case specific transformations. \citet{skilling2006nested} also mentions this issue and proposes to add a uniform random variable on a \emph{tiny} interval but one lacks of justifications and clear guidelines and consequences.

Following the random walk framework from \citep{walter2014point} the goal of this paper is to fill this gap by providing both the distribution of the number of iterations and the MVUE estimator in each case: the one with strict inequality and the one with non-strict inequality. Indeed the decomposition of \eqref{eq:definiton mutlilevel splitting} suggests these two possible definitions for a Multilevel Splitting method. While there are the same with probability $1$ if $X$ is continuous, they may differ if $X$ is not. Especially recent results from \cite{brehier2015unbiasedness} derive an unbiased estimator for the strict inequality case in the usual AMS framework. However the distributions of the estimators are less simple than in the continuous case. In this scope we also suggest a third estimator based on the random walk with non-strict inequality which has the same statistical properties as in the continuous case. Practically speaking it is not necessary to know in advance if $X$ is actually continuous or not and in this latter case, the three estimators become the same.

The paper is organized as follows: first we recall the theoretical results for continuous random variables. Then we show how discontinuities alter them and we give corresponding \emph{corrected} estimators. Finally, we apply this modified version to a dynamic test case from \citep{simonnet2014combinatorial}, it is the probability of a Markov process reaching a given set $B$ before going to another set $A$, and to a counting problem, namely the \emph{Boolean SATisfiability problem} \citep{hoos2000satlib}.

\section{Rare event simulation for discontinuous random variables}

\subsection{The increasing random walk}
In this section, we recall common results from \citep{huber2011using,guyader2011simulation,simonnet2014combinatorial} in the framework of \citep{walter2014point}. Let us consider $X = S(\U) \in \R$ a real-valued random variable with distribution $\mu^X$ where $S$ is a deterministic function (for instance the output of a computer code) and $\U$ a random finite- or infinite-dimensional vector with known distribution $\mu^U$. In this section we assume that the \cdf $F_X$ of $X$ is continuous.

\begin{dfntn}[Increasing random walk] \label{def:increasing random walk}
Let $X$ be a real-valued random variable with continuous \cdf $F_X$, $X_0 = -\infty$ and define the Markov sequence $(X_n)_n$ such that:
\begin{equation} \label{eq:def increasing random walk}
\forall n \in \N,\; \proba{X_{n+1} \in A \mid X_0, \cdots, X_n} = \dfrac{\proba{X \in A \cap (X_n, +\infty)}}{\proba{X \in (X_n, +\infty)}}
\end{equation}
\end{dfntn}
In other words $(X_n)_n$ is an increasing sequence where each element is randomly generated conditionally greater than the previous one: $X_{n+1} \sim \mu^X( \cdot \mid X > X_n)$. Assuming that $F_X$ is continuous, the associated sequence $(T_n)_n$ such that $T_n = - \log \proba{X > X_n} $ is distributed as the arrival times of a Poisson Process with parameter 1.
\begin{rmrk} \label{rem:equivalence strict nstrict continu}
Since $X$ is continuous, the random walk can alternatively be defined with non-strict inequality: $X_{n+1} \sim \mu^X( \cdot \mid X \geq X_n)$ without changing this result.
\end{rmrk}

\begin{figure}[!ht]
\centering
\def\svgwidth{0.75\textwidth}
\hspace{-40pt}
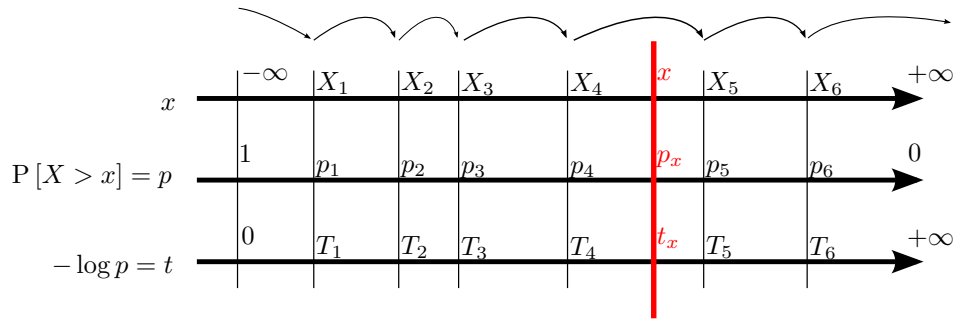
\caption{The increasing random walk and the associated Poisson process}
\label{fig:schema APDP}
\end{figure}

In the sequel, we then consider for all $x \in \R$ the associated numbers $p_x = \proba{X > x}$ and $t_x = -\log p_x$. Hence, for all $x \in \R$, the counting random variable of the number of events before $x$: $M_x = \operatorname{card}\left\lbrace n \in \N \mid X_n \leq x \right\rbrace$ follows a Poisson law with parameter $t_x = -\log p_x$. From this result one can build an estimator for $p_x$. Indeed, simulating $N$ \iid Poisson random variables with the same parameter $t_x$: $(M_x^i)_{i = 1}^N$, one seeks for estimating $e^{-t_x}$. In this context the Lehmann-Scheffé theorem insures that the Minimal-Variance Unbiased Estimator (MVUE) is:
\begin{equation} \label{eq:definition proba estimator}
\p_x = \l 1 - \dfrac{1}{N} \r^{\Sum[i=1][N] M_x^i}
\end{equation}
\begin{prpstn}[Statistical properties of the probability estimator] \label{propo:statistical properties p_x}
\begin{align}
\p_x &\xrightarrow[N \to \infty]{a.s} p_x \\
\var{\p_x} &= p_x^2 \l p_x^{-1/N} - 1 \r \label{eq:var estimator continuous}
\end{align}
\end{prpstn}
This estimator exhibits a logarithmic efficiency and asymptotically achieves the Cramer-Rao bound $-p_x^2 \log p_x / N$.
\begin{rmrk}
The Last Particle Algorithm \citep{guyader2011simulation,simonnet2014combinatorial} is only one possible implementation of this estimator; especially \citep{walter2015moving} studied its parallelisation and showed that it generates a marked Poisson process with parameter $N$. On the other hand, \citep{huber2011using} presented the increasing random walk linked with a Poisson process with parameter 1 but focused on $\log p_x$ instead of $p_x$.
\end{rmrk}

\begin{rmrk}
When generating $M_x$ a counting random variable at given \emph{state} $x \in \R$, one indeed generates $M_{x_0}$ for all $x_0 \leq x$. This gives a Glivenko-Cantelli like results:
\begin{equation} \label{eq:glivenko cantelli}
\forall x_0 \leq x,\; F_N(x_0) = 1 - \l 1 - \dfrac{1}{N} \r^{M_{x_0}} \xrightarrow[N \to \infty]{a.s.} F_X(x_0)
\end{equation}
with $M_{x_0}$ the sum of $N$ \iid counting random variables at \emph{state} $x_0$.
\end{rmrk}
With several realisations of the increasing random walk it is also possible to define parallel quantile and mean estimators for continuous random variables. The interested reader is referred to \citep{guyader2011simulation} and \citep{walter2015moving} for quantile estimation and \citep{walter2014point} for mean estimation.

\subsection{Extension of the random walk for discontinuous random variables}
We now intend to extend the definition of the random walk of Section \ref{ss:The increasing random walk} to the case where $X$ is not necessarily continuous -- or potentially discrete. Especially, while for all $x \in \R$, $\proba{X > x} = \proba{X \geq x}$ when $F_X$ is continuous, there are now two alternative definitions for the increasing random walk (see Definition \ref{def:increasing random walk}), precisely the one with strict inequality and the one with non-strict inequality (see Remark \ref{rem:equivalence strict nstrict continu}). In this latter case \citet{simonnet2014combinatorial} showed that the counting random variable at a given \emph{state} $x$ follows indeed a mixture of a Poisson law and independent Geometric laws with parameters depending on the atoms of $X$. This result prevented him from getting an unbiased estimator for $p_x$. Instead he focused on the special case with only one jump point $x_0$ reached at the first iteration of the random walk. In this context he proposed to estimate the atom with a crude Monte Carlo on the first \iid sampling and then to consider the continuous random variable $X$ with the conditional distribution given $X > x_0$, that is the distribution with \cdf $\one_{x>x_0} \l F_X(x) - F_X(x_0) \r / \l 1 - F_X(x_0) \r$ \citep[Theorem 5]{simonnet2014combinatorial}.

In this paper we go one step beyond his work by providing a general MVUE for the non-strict inequality random walk as well as for the strict inequality random walk. We also introduce a method for getting an estimator with the same properties as above (Proposition \ref{propo:statistical properties p_x}) even if the random variable is discontinuous, but without any intrusive nor case specific trick as in \citep{cerou2011use} or \citep{huber2011using}.

From now on, we assume that $X$ is a real-valued random variable, continuous or not, and we denote by $D$ the set of its atoms. According to Froda's theorem \citep{froda1929distribution} it is countable. As in Section \ref{ss:The increasing random walk} we consider for all $x \in \R$ the counting random variable $M_x = \operatorname{card} \{ n \in \N \mid X_n \leq x \}$. The next two propositions aim at giving the law of $M_x$ in the strict and non-strict cases. In both Propositions \ref{propo:loi M non strict} and \ref{propo:loi M strict}, we consider $x \in \R \mid p_x = \proba{X > x} > 0$, $D_x = D \cap (-\infty, x]$ and define: $\forall d \in D_x$:
\begin{equation} \label{eq:def delta}
\Delta_d = \dfrac{\proba{X > d}}{\proba{X \geq d}}.
\end{equation}

\begin{prpstn}[Law of the counting random variable for the non-strict random walk] \label{propo:loi M non strict}
\begin{equation} \label{eq:loi M non strict}
M_x \sim \mathcal{P}\l -\log \dfrac{p_x}{\prod \limits_{d \in D_x} \Delta_d } \r + \Sum[d \in D_x][] \mathcal{G}\l \Delta_d \r
\end{equation}
with $\mathcal{G}$ a Geometric law counting the number of failures before success.
\begin{proofprop}
The distribution of $M_x$ has already been proved by \citep{simonnet2014combinatorial} assuming that $\card D_x < \infty$. We extend this result to the possible infinite countable number of discontinuity.

Let $S_n = \{ x \in \R \mid 0 < \proba{X = x} < 1/n \}$ be the set of the jump points of $F_X$ with jump amplitudes smaller than $1/n$ and $X^{(n)} = X \one_{X \notin S_n}$. $X^{(n)}$ has a finite number of jump points and accumulates the (possibly infinite) number of jump points $d \in D \mid \proba{X = d} < 1/n$ at $0$. Since it has a finite number of discontinuities, the law of its associated counting random variable is then known.

Furthermore it verifies:
\[
\forall x \in \R,\; \proba{X^{(n)} \leq x} = \proba{X \leq x} + \one_{x \geq 0} \Sum[\substack{d \in S_n \\ d > x}][] \proba{X = d} - \one_{x<0} \Sum[\substack{d \in S_n \\ d \leq x}][] \proba{X = d}.
\]
Hence $X^{(n)} \xrightarrow[n \to \infty]{\mathcal{L}} X$, which implies $M_x^{(n)} \xrightarrow[n \to \infty]{\mathcal{L}} M_x$ with $M_x^{(n)}$ the counting random variable at state $x$ associated with an increasing random walk on $X^{(n)}$.

Moreover, one has: $\forall x \in \R,\; \forall n \geq 1,\; M_x^{(n)} \sim \mathcal{P} \l -\log \dfrac{\proba{X^{(n)} > x}}{\prod \limits_{d \in D_x \setminus S_n} \Delta_d} \r + \Sum[d \in D_x \setminus S_n][] \mathcal{G}\l \Delta_d \r$. Finally, this gives:
\[
M_x \sim \mathcal{P} \l -\log \dfrac{\proba{X > x}}{\prod \limits_{d \in D_x} \Delta_d} \r + \Sum[d \in D_x][] \mathcal{G}\l \Delta_d \r .
\]
We also show that the first and second order moments of $M_x$ remain finite even when $\operatorname{card}(D_x) = \infty$. One has:
\begin{align*}
0 \leq \Sum[d \in D_x][] \l \dfrac{1}{\Delta_d} - 1 \r = \Sum[d \in D_x][] \dfrac{\proba{X = d}}{\proba{X > d}} 
\leq \dfrac{1}{\proba{X > x}} \Sum[d \in D_x][] \proba{X = d} \leq \dfrac{1-p_x}{p_x},
\end{align*}
and:
\[
1 \geq \prod \limits_{d \in D_x} \Delta_d \geq \prod \limits_{d \in D_x} e^{-\l \frac{1}{\Delta_d} - 1 \r} \geq e^{-\Sum[d \in D_x][] \l \frac{1}{\Delta_D} - 1 \r} \geq e^{-\frac{1-p_x}{p_x}} > 0.
\]
All together, theses inequalities give the result:
\begin{align*}
\E{M_x} &= -\log \dfrac{p_x}{\prod \limits_{d \in D_x} \Delta} + \Sum[d \in D_x] \l \dfrac{1}{\Delta_d} - 1 \r \leq -\log p_x + \dfrac{1 - p_x}{p_x} \\
\var{M_x} &= -\log \dfrac{p_x}{\prod \limits_{d \in D_x} \Delta} + \Sum[d \in D_x] \dfrac{1}{\Delta_d} \l \dfrac{1}{\Delta_d} - 1 \r \leq -\log p_x + \dfrac{1 - p_x}{p_x^2} .\\
\end{align*}
\end{proofprop}
\end{prpstn}

The distribution of $M_x$ is that of the sum of a Poisson random variable with parameter $-\log p_x / \prod \Delta_d$ and of Geometric random variables with parameters $(\Delta_d)_{d \in D_x}$; these random variables being independent.  It is part of the proof given in Appendix that the distribution is well defined with finite mean and variance even when $\card(D_x) = \infty$. Also this result was already stated in \citep{simonnet2014combinatorial} with a combinatorial analysis assuming that $\operatorname{card}(D) < \infty$. Indeed it can be understood using the renewal property of a Poisson Process: the number of events corresponding to the \emph{continuous} part, \ie{} events $X_n \notin D$, follows a Poisson law with parameter $-\log p_x - \sum (-\log \Delta_d) = -\log (p_x / \prod \Delta_d)$. On the other hand each jump point leads to a random number of iterations following a Geometric law with probability of success $\proba{X > d}/\proba{X \geq d} = \Delta_d$.

\begin{prpstn}[Law of the counting random variable for the strict random walk] \label{propo:loi M strict}
\begin{equation} \label{eq:loi M strict}
M_x \sim \mathcal{P}\l -\log \dfrac{p_x}{\prod \limits_{d \in D_x} \Delta_d } \r + \Sum[d \in D_x][] \mathcal{B}\l 1 - \Delta_d \r
\end{equation}
with $\mathcal{B}$ a Bernoulli distribution.
\begin{proofprop}
Using the renewal property of the Poisson Process the number of events in the \emph{continuous} part will be the same as the one in the non-strict case. Indeed the only difference with the non-strict random walk comes from the \emph{behaviour} of the random walk when $X_n \in D_x$. In this latter case, while the non-strict inequality repeats the trial until success, the strict inequality do it only once. Hence the Geometric law is replaced by a Bernoulli one.
\end{proofprop}
\end{prpstn}

Hence the strict inequality random walk \emph{replaces} the Geometric distribution by a Bernoulli one. Since the Geometric law counts the number of failures while the Bernoulli one gives $1$ in case of success, the parameter is the opposite. Furthermore, both \eqref{eq:loi M non strict} and \eqref{eq:loi M strict} are equal when $D = \varnothing$, \ie{} when $X$ is continuous. In this latter case, one finds back the pure Poisson distribution $M_x \sim \mathcal{P}(-\log p_x)$.

\subsection{Probability estimators}
As noticed by \citep{simonnet2014combinatorial} the formulas \originaleqref{eq:loi M non strict}-\originaleqref{eq:loi M strict} are not very useful to derive an unbiased probability estimator. However we do not generate only \iid copies of the counting random variables but the random walks themselves. Hence if one can afford storing all the \emph{states} of each random walk, then it is possible to build a MVUE in both cases.

\begin{lmm}[MVUE for a Geometric distribution] \label{lem:MVUE Geom}
Let $G \sim \mathcal{G}(p)$ be a Geometric random variable counting the number of failures before success with probability of success $p$ and $(G_i)_{i=1}^N$ $N$ \iid copies of $G$, then the minimal variance unbiased estimator for $p$ is:
\begin{equation}
\p = \dfrac{N-1}{N-1 + \Sum G_i}
\end{equation}
\begin{prooflem}
One is going to use Lehmann-Scheffé theorem with the statistic $T = \Sum G_i$. As the sum of $N$ independent Geometric random variables with parameter $p$, $T$ follows a Negative Binomial law: $\forall t \in \N, \; \proba{T = t} = \binom{N+t-1}{t} p^N (1-p)^t$. $T$ is sufficient:
\[
\mathcal{L}(g_1, \cdots, g_N, p) = \prod \limits_{i = 1}^{N} \proba{G_i = g_i} = (1-p)^{\Sum g_i} p^N = (1-p)^t p^N .
\]
$T$ is also complete: let $\phi$ be a measurable function, one has:
\begin{align*}
\forall p \in (0,1),\; \E{\phi(T)} = 0 &\Rightarrow \forall p \in (0,1),\; p^N \Sum[t = 0][\infty] \binom{t + N -1}{t} \phi(t) (1-p)^t = 0 \\
&\Rightarrow \forall \theta \in (0,1),\; \Sum[t = 0][\infty] \alpha_t \theta^t
\end{align*}
with $\alpha_t = \binom{t + N - 1}{t} \phi(t)$ and $\theta = 1-p$. Furthermore $p=1$ \ie{} $\theta = 0$ gives $\phi(0) = 0$ and $\theta = 1$, \ie{} $p = 0$ gives $\proba{T < \infty} = 0$. Hence the power series $\theta \mapsto \sum \alpha_t \theta^t$ is identically null on its radius of convergence $[0, 1)$ and so $\forall t \in \N, \; \alpha_t = 0$, which means $\forall t \in \N,\; \phi(t) = 0$ and $T$ is complete.

We now consider the estimator $R = \dfrac{1}{N} \Sum \one_{G_i=0}$. $R$ is unbiased because $\E{\one_{G_i=0}} = \proba{G_1 = 0} = p$. Then Lehmann-Scheffé theorem states that $\E{R \mid T}$ is the MVUE of $p$. This gives:
\begin{align*}
\E{R \mid T = t} &= \dfrac{1}{N} \Sum \E{ \one_{G_i=0} \mid \Sum G_i = t} = \proba{G_1 = 0 \mid \Sum G_i = t} \\
&= \dfrac{\proba{G_1 = 0, \Sum[i=2][N] G_i = t}}{\proba{\Sum G_i = t}} = \proba{G_1 = 0} \dfrac{\binom{t + N - 2}{t} p^{N-1} (1-p)^t}{\binom{t + N - 1}{t} p^{N} (1-p)^t} \\
\E{R \mid T = t} &= \dfrac{N-1}{N-1 + t}
\end{align*}
Hence, $\p = \E{R \mid T} = (N-1)/(N-1 + T)$ is the MVUE of $p$.
\end{prooflem}
\end{lmm}

\begin{lmm}[MVUE for a Bernoulli distribution] \label{lem:MVUE bernoulli}
Let $B \sim \mathcal{B}(1-p)$ be a Bernoulli random variable with probability of failure $p$ and $(B_i)_{i=1}^N$ $N$ \iid copies of $B$, then the minimal variance unbiased estimator for $p$ is:
\begin{equation}
\p =  1 - \dfrac{\Sum B_i}{N}
\end{equation}
\end{lmm}

\begin{dfntn}[Run-length encoding]
Let $\mathbf{v} = (v_1, \cdots, v_m) \in \R^m,\; m \geq 1$ be a vector such that $\forall i \in \llbracket 1, m-1 \rrbracket,\; v_i \leq v_{i+1}$. We call the \emph{run-length encoding} of $\mathbf{v}$ the vector $\mathbf{r}$ of the lengths of runs of equal values in $\mathbf{v}$.
\end{dfntn}
In other words, the run-length encoding counts for any non decreasing sequence the number of times each value is repeated: for example if $\mathbf{v} = (0.5,2.1,2.1,2.1,\pi)$ then $\mathbf{r} = (1, 3, 1)$. Especially,  if $X$ is continuous the RLE of the states of a realisation of the increasing random walk $(X_1, \cdots, X_m)$, is $\mathbf{r} = (1, \cdots, 1) \in \R^m$ with probability 1 while on the contrary discontinuities will produce repeated values with non-zero probability. More precisely, the number of times each value is repeated corresponds to the number of failures while sampling above a threshold. With this consideration we are now in position to define the probability estimators.

In the sequel we assume that for a given $x \in \R \mid \proba{X > x} > 0$, $(X_i)_{i=1}^{M_x}$ is the merged and sorted sequence of the states of $N$ (non-)strict inequality random walks generated until state $x$; $M_x = \Sum M_x^i$ is the sum of the counting random variables of each random walk, $\mathbf{r}$ is the RLE of $(X_1, \cdots, X_{M_x})$, and $l$ is its length.

\begin{prpstn} \label{propo:MVUE non strict random walk}
The MVUE for the non-strict inequality random walk is:
\begin{equation} \label{eq:MVUE non strict random walk}
\p_x = \prod \limits_{i = 1}^{l} \dfrac{N-1}{N-1 + r_i} .
\end{equation}
It verifies:
\begin{equation} \label{eq:variance MVUE non strict}
p_x^2 \l p_\text{pois}^{-1/N} - 1 \r \leq \var{\p_x} \leq p_x^2 \l p_\text{pois}^{-1/N} \l \dfrac{N-1}{N-2} \r^{\# D_x} - 1 \r
\end{equation}
with $p_\text{pois} = \dfrac{p_x}{\prod \limits_{d \in D_x} \Delta_d}$ and $\# D_x = \operatorname{card}(D_x)$.
\begin{proofprop}
On the one hand, for all $a<b$ such that $X$ is continuous on $(a, b)$, $\proba{X > b \mid X > a}$ can be estimated by $(1-1/N)^{\# \{ X_n \in (a,b) \}}$ with $\# \{ X_n \in (a,b) \}$ the number of events of the sum process in $(a,b)$. Moreover, since $b \mapsto \proba{X \geq b}$ is left-continuous, it is also an MVUE of $\proba{X \geq b \mid X > a}$ and this relation remains true if $b \to a$ since $\operatorname{card}(\varnothing) = 0$. Using the fact that the RLE of $(X_1, \cdots, X_{M_x})$, $\mathbf{r} = (1, ..., 1)$ with probability $1$ when $X$ is continuous, $(1-1/N)^{\# \{ X_n \in (a,b) \}} = \prod (N-1)/(N-1 + r_i)$ with probability $1$.

On the other hand $\forall d \in D_x, \; \Delta_d = \proba{X > d}/\proba{X \geq d}$ can be estimated with the MVUE defined in Lemma \ref{lem:MVUE Geom}. The first state of each chain non smaller than $d$ can be considered as an \iid sample of $X \mid X \geq d$. The number of times $d$ is found in $(X_1, \cdots, X_{M_x})$ is then the sum of $N$ realisations of a Geometric random variable with parameter $\Delta_d$. Hence, $\Delta_d$ is estimated with $(N-1)/(N-1+r_d)$ with $r_d$ the number of times $d$ is found in $(X_1, \cdots, X_{M_x})$.

Since $D$ is countable, we consider $\R \setminus D = \bigcup_i \, I_i$ with $(I_i)_i$ a sequence of disjoint open intervals. Note that some subsequence of $I_n$ may converge toward the empty set if $D$ is infinite countable with some accumulation points. However $X$ is continuous on $\R \setminus D$. Using the renewal property of the Poisson process, one can consider that $\prod (N-1)/(N-1 + r_i)$ is a product of independent MVUE estimators. Especially, denoting by $M_\text{pois}$ the number of $1$ in $\mathbf{r}$, $(1-1/N)^{M_\text{pois}}$ is a MVUE of $p_\text{pois} := p_x / \prod \Delta_d$.

We now explicit the calculation for the bounds on the variance. One has:
\begin{align*}
\E{\p_x^2} = p_\text{pois}^2 p_\text{pois}^{-1/N} \prod \limits_{d \in D_x} \E{ \l \dfrac{N-1}{N-1 + T_d} \r^2}
\end{align*}
with $T_d \sim NegBin(N, \Delta_d)$. For a given $d \in D_x$, one has:
\begin{align*}
\E{\l \dfrac{N-1}{N-1 + T_d} \r^2} &= \Sum[t = 0][\infty] \binom{t + N - 1}{t} (1-\Delta_d)^t \Delta_d^N \l \dfrac{N-1}{N-1+t} \r^2 \\
&= \Sum[t = 0][\infty] \binom{t + N - 2}{t} (1-\Delta_d)^t \Delta_d^N \dfrac{N-1}{N-1+t} \\
&= \Delta_d \Sum[t = 0][\infty] \binom{t + N - 2}{t} (1-\Delta_d)^t \Delta_d^{N-1} \dfrac{N-2}{N-2+t} \dfrac{N-1}{N-2} \dfrac{N-2+t}{N-1+t}.
\end{align*}
Furthermore, $\forall t \geq 0,\; \dfrac{N-1}{N-2} \dfrac{N-2+t}{N-1+t} \in [1, (N-1)/(N-2)]$, which gives:
\[
\Delta_d^2 \leq \E{\l \dfrac{N-1}{N-1 + T_d} \r^2} \leq \Delta_d^2 \dfrac{N-1}{N-2}.
\]
Eventually the variance writes:
\begin{align*}
p_x^2 \l p_\text{pois}^{-1/N} - 1 \r \leq \var{\p_x} &= \E{\p_x^2} - p_x^2
%\leq p_\text{pois}^{2-1/N} \prod \limits_{d \in D_x} \Delta_d^2 \dfrac{N-1}{N-2} - p_x^2 = p_x^2 \l p_x^{-1/N} \prod \limits_{d \in D_x} \Delta_d^{1/N} \dfrac{N-1}{N-2} - 1 \r
\leq p_x^2 \l p_\text{pois}^{-1/N} \prod \limits_{d \in D_x} \dfrac{N-1}{N-2} - 1 \r .
\end{align*}
\end{proofprop}
\end{prpstn}

It is interesting to notice here that in a case of a discrete random variable, $p_\text{pois} = 1$ and the bounds on the variance become:
\begin{equation} \label{eq:bounds variance nstrict discret}
0 \leq \dfrac{\var{\p_x}}{p_x^2} \leq \l \dfrac{N-1}{N-2} \r^{\# D_x} - 1 .
\end{equation}
It means that the coefficient of variation is bounded by a quantity which does not depend on the size but only on the number of jumps. Since this quantity is likely to be known with good confidence and does not vary much with $\# D_x$, this can provide a robust upper bound for the coefficient of variation.
\begin{prpstn} \label{propo:MVUE strict random walk}
The MVUE for the strict inequality random walk is:
\begin{equation} \label{eq:MVUE strict random walk}
\p_x = \prod \limits_{i = 1}^{l} \l 1 - \dfrac{r_i}{N} \r .
\end{equation}
It verifies:
\begin{equation} \label{eq:variance MVUE strict}
\var{\p_x} = p_x^2 \l p_x^{-1/N} \prod \limits_{d \in D_x} g(\Delta_d, N) - 1 \r
\end{equation}
with $g: (\Delta, N) \mapsto \dfrac{\Delta (N-1) + 1}{N \Delta^{1-1/N}}$.
\begin{proofprop}
The same reasoning	as for the proof of Proposition \ref{propo:MVUE non strict random walk} applies where the MVUE of a Geometric law is replaced by the one of a Bernoulli distribution. For the variance one has:
\begin{align*}
\E{\p_x^2} &= p_\text{pois}^2 \; p_\text{pois}^{-1/N} \prod \limits_{d \in D_x} \dfrac{\Delta_d^2}{N} \l N-1 + \dfrac{1}{\Delta_d} \r \\
\E{\p_x^2} &= p_x^2 \; p_x^{-1/N} \prod \limits_{d \in D_x} \dfrac{\Delta_d (N-1) + 1}{N \Delta_d^{1-1/N}}
\end{align*}
which gives the result.
\end{proofprop}
\end{prpstn}

On the one hand we have been able to define minimal variance unbiased estimators for both the strict and non-strict random walks. They become equal when the random variable is continuous and in this case one finds back the estimator defined in Section \ref{ss:The increasing random walk}. Especially the variance increase due to discontinuities in the distribution of $X$ is clearly visible in \eqref{eq:variance MVUE strict} as $g > 1$ if $\Delta < 1$. On the other hand their distributions are not easy to characterise; in particular we could not derive any practical literal expression of the variance in the non-strict case. In this context we suggest to consider an auxiliary continuous random variable which involves an independent uniform random variable. This transformation is general and does not require any other knowledge on the problem.

\begin{figure}
\begin{knitrout}
\definecolor{shadecolor}{rgb}{0.969, 0.969, 0.969}\color{fgcolor}

{\centering \includegraphics[width=\maxwidth]{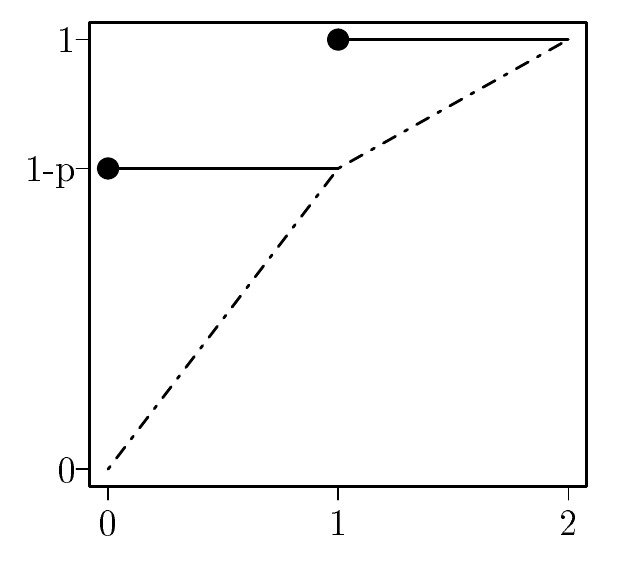} 

}

\end{knitrout}
\caption{\cdf of a Bernoulli random variable $B \sim \mathcal{B}(p)$ (plain dark line) and its associated continuous random variable $Y = B + U$ (dark dashed line) with $U \sim \mathcal{U}[0, 1]$.}
\label{fig:transformation cdf}
\end{figure}

Indeed when generating a Geometric random variable with \iid Bernoulli trials, one can also consider the random variable $Y = B + U$ with $B \sim \mathcal{B}(p)$ the Bernoulli random variable with probability of success $p$ and $U \sim \mathcal{U}[0, 1]$ an independent Uniform random variable on the interval $[0, 1]$. Figure \ref{fig:transformation cdf} plots the \cdf of $B$ and $Y$. Furthermore, $\forall y \in [0, 2], \{Y > y \} = \left\lbrace \{ B \geq \lfloor y \rfloor \} \cap \{U > Y - \lfloor y \rfloor \} \right\rbrace$. Practically speaking, this means that the generation of the geometric random variable can be seen as a basic Acceptance-Rejection scheme used to generate the increasing random walk on $Y$ until state $1$: for each generated $B$, sample also $U \sim \mathcal{U}[0,1]$ and accept the transition for $Y$ if $U > y - \lfloor y \rfloor$. Eventually, considering the fact that $p = \proba{B = 1} = \proba{Y \geq 1} = \proba{Y > 1}$, $p$ can also be estimated using the increasing random walk on the continuous random variable $Y$. To conclude, in addition to the MVUE of $p$ defined in Lemma \ref{lem:MVUE Geom} and at the cost of the generation of an independent uniform random variable, one also produces an estimator of the form of \eqref{eq:definition proba estimator} with the same statistical properties. Embedding this in the generation of the non-strict inequality random walk gives then an estimator with the same properties as the one in the continuous case. Algorithm \ref{algo:pseudo code Y}, Theorem \ref{th:generation M Y} and Corollary \ref{coro:pure poisson estimator} precise this point. In the sequel, we will refer to this estimator as the \emph{pure Poisson} estimator.

\begin{algorithm}[!ht]
\caption{Pseudo-code for the non-strict inequality random walk and the pure Poisson estimator}
\label{algo:pseudo code Y}
\begin{algorithmic}[3]
\Require $x$
\State $M = 0$
\State Draw $X_1 \sim \mu^X$ and $U_1 \sim U[0; 1]$; $n = 1$
\While{$X_n \leq x$}
\State $M = M + 1$
\State Draw $X_{n+1} \sim \mu^X( \cdot \mid X \geq X_n)$ and $U_{n+1} \sim U[0; 1]$
\While{$X_{n+1} = X_n$} \label{line:debut while test X in D}
\If{$U_{n+1} > U_n$} \label{line:AR section algo sur Y}
\State $M = M + 1$
\EndIf
\State $n = n + 1$
\State Draw $X_{n+1} \sim \mu^X( \cdot \mid X \geq X_n)$ and $U_{n+1} \sim U[0; 1]$
\EndWhile \label{line:fin while test X in D}
\State $n = n + 1$	
\EndWhile
\State \textbf{return} $M,\; (X_n)_n$
\end{algorithmic}
\end{algorithm}

\begin{thrm} \label{th:generation M Y}
In Algorithm \ref{algo:pseudo code Y}, the random variable $M$ follows a Poisson law with parameter $-\log \proba{X > x}$.
\begin{prooftheo}
The difference between the generation of the non-strict random walk and Algorithm \ref{algo:pseudo code Y} stands in the addition of the \emph{while} loop from line \ref{line:debut while test X in D} to line \ref{line:fin while test X in D}. This loop is entered when a Geometric scheme is started: two consecutive events being equal is a non-zero probability event only when $X_n \in D$. In this context, the condition $\{U_{n+1} > U_n \}$ lets generate the counting random variable of the continuous random variable associated with the Geometric scheme as explained in Section \ref{ss:Probability estimators}. Therefore it follows a Poisson distribution with parameter $- \log \Delta_{X_n}$. The renewal properties of the Poisson Process lets conclude the proof. 
\end{prooftheo}
\end{thrm}

\begin{crllr} \label{coro:pure poisson estimator}
Let $N \geq 2$ and $M = \sum_{i=1}^N M_i$ the sum of $N$ \iid realisations of Algorithm \ref{algo:pseudo code Y}, the estimator
\begin{equation} \label{eq:pure poisson estimator}
\p_x = \l 1 - \dfrac{1}{N} \r^M
\end{equation}
has the same properties as in Proposition \ref{propo:statistical properties p_x}.
\end{crllr}

Finally the distinction between the strict and non-strict random walks lets define two different estimators for the probability of exceeding a threshold. Both are unbiased and become the same when $X$ is indeed continuous. However their distributions are not well-characterised. In this scope we have introduced a third estimator based on the non-strict random walk. With the addition of a \emph{while} loop and a Uniform random variable, we have been able to produce an estimator which has always the same statistical properties, $X$ being continuous or not. This estimator is not optimal in terms of variance when $X$ is actually discontinuous but remains close to it when the jumps $(\Delta_d)_d$ remain close to $1$: the MVUE of Lemma \ref{lem:MVUE Geom} has a squared coefficient of variation approximately equal to $(1-\Delta)/N$ while it is $-\log(\Delta)/N$ for the \emph{pure Poisson} estimator. These results are illustrated in the next Section.

\section{Examples}

\subsection{Discontinuous random variable}

\subsubsection*{Problem setting}
We consider here the example used by \citet{simonnet2014combinatorial} to illustrate his results. It is a numerical study with a Euler scheme of a diffusive process satisfying:
\begin{equation}
d \U_t = - \nabla V dt + \sqrt{\dfrac{2}{\beta}} d \mathbf{W}_t,\; \U_0 = \u_0
\end{equation}
where $\mathbf{W}_t$ is a Wiener process, $\beta^{-1}$ is the temperature, and $V$ is a potential defined by:
\[
V(u_1, u_2) = - \l \dfrac{u_1^2}{2} - \dfrac{u_1^4}{4} \r - b \l \dfrac{u_2^2}{2} - \dfrac{u_2^4}{4} \r + \dfrac{a}{2} u_1^2 u_2^2 ,
\]
for some $a$ and $b$. The goal is to estimate the probability that the process enters a given set $B$ before another set $A$ from an initial state $\u_0$: if $\tau_C$ is the stopping time defined by $\tau_C := \inf \left\lbrace t \geq 0 \mid \U_t \in C \right\rbrace$, then one seeks for estimating:
\[
p = P_{\u_0}[\tau_B < \tau_A],
\]
where $P_{\u_0}$ is the distribution of $(\U_t)_t$ starting from $\U_0 = \u_0$. As a function of $\u_0$ this quantity is known as the \emph{Committor}. From a practical point of view, it is often intractable and a \emph{reaction coordinate} is introduced to measure \emph{how far} a trajectory is escaping from $A$ before returning to it: $\Phi : \R^d \to \R$ such that $A = \Phi^{-1}\l (-\infty, 0] \r$ and $B = \Phi^{-1}\l [1, +\infty) \r$. With these notations, a trajectory enters $B$ before returning to $A$ if and only if:
\[
X := \underset{t \in [0, \tau_A)}{\sup} \Phi(\U_t) \geq 1.
\]
$X$ is then the real-valued random variable of interest and the problem is indeed to estimate $\proba{X \geq 1}$. With this notation, the theoretical results of Section \ref{s:Rare event simulation for discontinuous random variables} can be used directly.

\subsubsection*{Conditional sampling}
This is the only requirement of the estimators defined in Section \ref{ss:Probability estimators} but also the crucial point. To distinguish between theoretical results on the estimator and possible issues due to imperfect conditional sampling, \citet{simonnet2014combinatorial} presents two ways of generating conditional random variables, referred to as the \emph{Perfect} and the \emph{Effective} implementations. In the \emph{Perfect} case, Acceptance-Rejection sampling is used to generate samples above a given threshold while in the \emph{Effective} one, trajectories already following the targeted distributions are used to find an initial state $\u$ such that $\Phi(\u)$ is greater or equal than the current threshold. In this example, we want to focus on the theoretical results on the number of iterations and on the estimators. Hence we make use of the \emph{Perfect} sampling described in Algorithms \ref{algo:perfect sampling}.
\begin{algorithm}
\caption{Perfect sampling of $\X \sim \mu^X( \cdot \mid X \geq x)$ for the diffusive process}
\label{algo:perfect sampling}
\begin{algorithmic}
\Require $x \in \R$
\Comment the current threshold one seeks to sample above
\State $X^* = -\infty$
\While{$X^* < x$}
\State Generate a new trajectory $\U_t$ starting from $\u_0$
\State $X^* = \underset{t \in [0, \tau_{A \cup B}]}{\sup} \Phi(U_t)$
\EndWhile
\end{algorithmic}
\end{algorithm}

%\begin{algorithm}
%\caption{Effective sampling of $\X \sim \mu^X( \cdot \mid X \geq x)$ for the diffusive process}
%\label{algo:effective sampling}
%\begin{algorithmic}
%\Require $(X_i)_{i=1..k}$ and $(\U_t^i)_{i=1..k}$
%\Comment $k$ is a number of random walks (see Definition \ref{def:increasing random walk})
%%; $(\U_t^i)_i$ and $(X_i)_i$ are the corresponding trajectories and realisations of the real-valued random variable
%\State $i^* = \argmin_i X_i$
%\State Pick $I$ randomly in $\llbracket 1, k \rrbracket \setminus \{i^*\}$
%\State Find $\tau = \underset{t}{\inf} \{ \Phi(U^I_t) \geq X_{i^*} \}$
%\State Generate a new trajectory $\U_t^{i^*}$ starting from $\U_{\tau}^I$
%\State $X_{i^*} = \underset{t \in [0, \tau_{A \cup B}]}{\sup} \Phi(U_t^{i^*})$
%\end{algorithmic}
%\end{algorithm}
%Note here that while Algorithm \ref{algo:perfect sampling} lets generate the random walks one after the other, or in parallel, Algorithm \ref{algo:effective sampling} requires to manipulate a batch of $k$ trajectories to get starting points above the current threshold. Subsequently they are generated sequentially. Reader is referred to \citep{walter2015moving} for more details on parallel implementation.
 
\subsubsection*{Numerical results}

Here we set $\Phi(\u) = 0.5(1 + u_1)$, $a = 0.6$, $b = 0.3$, $\u_0 = (\ensuremath{-0.9}, 0)$, $\beta = 10$ and $dt = 1$ as in \citep{simonnet2014combinatorial}. $\Phi(\u) = 0.5(1 + u_1)$ and so $\Phi(\u_0) = 0.05$: with a large timestep some trajectories will go directly into $A$, producing a discontinuity in the \cdf of $X$. We computed reference values using a crude Monte Carlo estimator with $N = \ensuremath{10^{6}}$ and found $p = 0.068$ and $\Delta = 0.396$. We then set $N = 300$ to get a coefficient of variation below $10\%$ because: $\cv{\p}^2 \approx -\log p / N \Rightarrow N \approx 268$.

We first focus on the distribution of the number of iterations described in Propositions \ref{propo:loi M non strict} and \ref{propo:loi M strict} and on the \emph{corrected} number of iterations to get a pure Poisson distribution (see Theorem \ref{th:generation M Y}).

\begin{figure}[!ht]
\centering
\subfloat[Strict random walk]{
\begin{knitrout}
\definecolor{shadecolor}{rgb}{0.969, 0.969, 0.969}\color{fgcolor}

{\centering \includegraphics[width=\maxwidth]{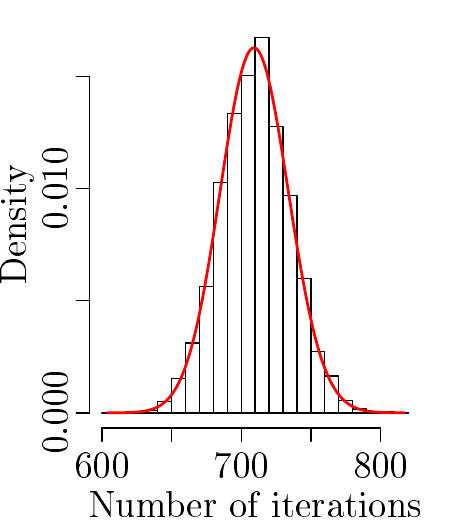} 

}

\end{knitrout}
\label{fig:niter strict}
}
\subfloat[Non-strict random walk]{
\begin{knitrout}
\definecolor{shadecolor}{rgb}{0.969, 0.969, 0.969}\color{fgcolor}

{\centering \includegraphics[width=\maxwidth]{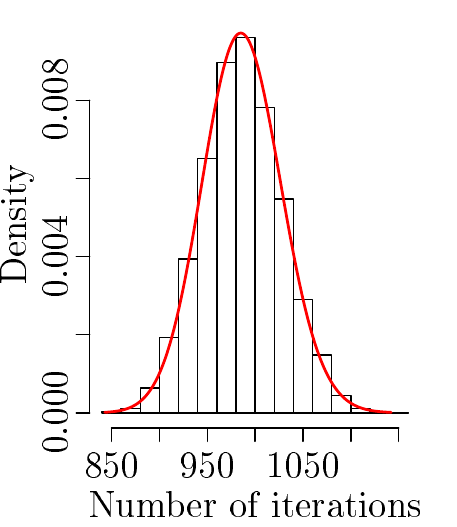} 

}

\end{knitrout}
\label{fig:niter nstrict}
}
\subfloat[Pure Poisson correction]{
\begin{knitrout}
\definecolor{shadecolor}{rgb}{0.969, 0.969, 0.969}\color{fgcolor}

{\centering \includegraphics[width=\maxwidth]{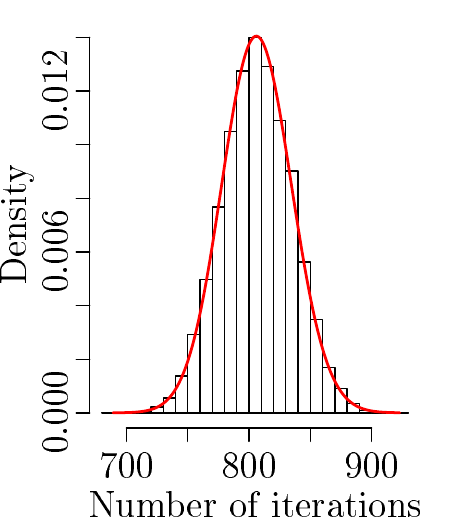} 

}

\end{knitrout}
\label{fig:niter poisson correction}
}
\caption{Histogram over $\ensuremath{10^{4}}$ realisations of the number of iterations for the strict random walk (\eqref{eq:loi M strict}), the non-strict one (\eqref{eq:loi M non strict}) and the pure Poisson \emph{correction} (Theorem \ref{th:generation M Y}). $X$ has one discontinuity at $x = 0.05$ and $\proba{X>x}/\proba{X\geq x} \approx 0.396$. The curves show the theoretical distributions.}
\label{fig:niter comp sim}
\end{figure}
Figure \ref{fig:niter comp sim} shows the histograms of the number of iterations for the strict random walk, the non-strict one and the pure Poisson \emph{correction}. Their are in good agreement with the theoretical distributions presented in \eqref{eq:loi M strict},  \eqref{eq:loi M non strict} and Theorem \ref{th:generation M Y} respectively. Especially we can see that for a given $N$, the cost of the estimators are different. Indeed, if one considers that the cost is the number of calls to a conditional simulator, then it is equal to the final number of iterations and Figure \ref{fig:niter strict} and \ref{fig:niter nstrict} present a clear shift: on average the discontinuity will produces $(1/\Delta - 1)N$ iterations for the non-strict random walk and only $(1-\Delta)N$ for the strict random walk; with $\Delta = 0.396$, this gives approximately $276$ more iterations. On the histograms a shift of $250$ to $300$ is clearly visible in the $x$ axis.

We now check the accuracy of the probability estimators. Firstly, they should be all unbiased. Secondly, for a given $N$ one should see a variance increase from the MVUE of the non-strict random walk of \eqref{eq:MVUE non strict random walk} to the pure Poisson estimator (\eqref{eq:pure poisson estimator}) and to the MVUE of the strict random walk (\eqref{eq:MVUE strict random walk}).

\begin{figure}[!ht]
\centering
\begin{knitrout}
\definecolor{shadecolor}{rgb}{0.969, 0.969, 0.969}\color{fgcolor}

{\centering \includegraphics[width=\maxwidth]{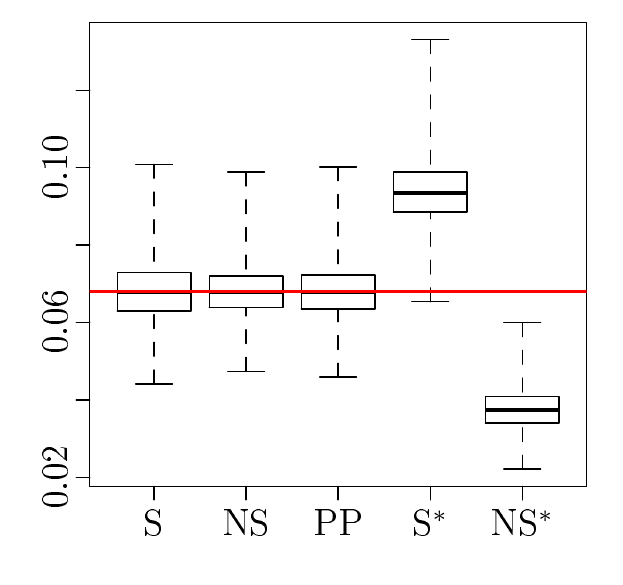} 

}

\end{knitrout}
\caption{Boxplots of the estimation of $p \approx 0.068$ over $\ensuremath{10^{4}}$ simulations with $N = 300$, whiskers extending to the extreme values. (S): Strict random walk estimator (\eqref{eq:MVUE strict random walk}), (NS): Non-strict random walk estimator (\eqref{eq:MVUE non strict random walk}), (PP): Pure Poisson estimator (\eqref{eq:pure poisson estimator}); with $^*$: estimators if the discontinuity is not taken into account, \ie{} $(1 - 1/N)^\text{Number of iterations}$.}
\label{fig:estimators sim}
\end{figure}

Figure \ref{fig:estimators sim} shows a boxplot of the three estimators over $\ensuremath{10^{4}}$ simulations. As an illustration, the estimators computed directly as if $X$ were continuous are also added to the plot. The horizontal line stands for the reference value calculated with a crude Monte Carlo. The estimated means are $0.068$ for the strict random walk estimator, $0.068$ for the non-strict one and $0.068$ for the pure Poisson one. This is in good agreement with the estimated reference value $p = 0.068$. Furthermore, the empirical variances are $\ensuremath{5.18\times 10^{-5}}$ for the strict inequality random walk and $\ensuremath{4.21\times 10^{-5}}$ for the pure Poisson estimator while the theoretical values given by \eqref{eq:variance MVUE strict} and \eqref{eq:var estimator continuous} are $\ensuremath{5.09\times 10^{-5}}$ and $\ensuremath{4.16\times 10^{-5}}$. On the other hand, the probability estimators are clearly not consistent when the discontinuity is not handled properly, \ie{} when the estimator is computed with the formula valid only in the continuous case (\eqref{eq:definition proba estimator}).

All together, these numerical results are in good agreement with the the theoretical ones. 

\subsection{Discrete random variable}

Counting problems are typical cases where the random variable of interest is known to be integer-valued. Indeed, it is shown that many of these problems can be put into the setting of estimating extreme probability \citep{mitzenmacher2005probability,bezakova2008accelerating,botev2008efficient,motwani2010randomized}. Among other, we focus here on the \emph{Boolean SATisfiability Problem} (\emph{SAT} problem). We do not pretend being competitive against specific \emph{SAT} solvers. Instead, we use this test case because the random variable will have several discontinuity points not only at the origin.

\subsubsection*{The \emph{SAT} problem} A SAT problem comprises
\begin{enumerate*}[label=\arabic*)]
\item a binary vector of $n$ \emph{literals} which can be either \texttt{TRUE} (=1) of \texttt{FALSE} (=0): $\u = (u_1, \cdots, u_n)$ is called a \emph{truth assignment}, \eg $\u = (\mathtt{TRUE}, \mathtt{TRUE}, \cdots, \mathtt{FALSE}) = (1, 1, \cdots, 0)$ and
\item a set of $m$ clauses $\{S_1, \cdots, S_m \}$ expressed as \texttt{OR} logical operators (also denoted by $\vee$) on the \emph{literals}, \eg $S_i = u_{i_1} \vee u_{i_2} \vee \cdots \vee u_{i_k}$.
\end{enumerate*}

The SAT problem in itself is then defined as follows: find an assignment $\u$ such that all clauses are true (\emph{SAT assignment problem}) or count the number of different assignments which satisfy all the clauses (\emph{Sharp-SAT}). The conjunctive normal form (CNF) of a SAT problem is then the product (logical operator \texttt{AND} or $\wedge$) of all clauses $F = S_1 \wedge \cdots \wedge S_m$ and both problems can be rewritten:
\begin{description}
\item[SAT assignment problem] is there at least one $\u \in \{0, 1\}^n$ such that $F(\u) = \mathtt{TRUE}$ ?
\item[sharp-SAT] find $card( \mathcal{S} ) = \, | \mathcal{S} | $ with $\mathcal{S} =  \{ \u \in \{0, 1\}^n \mid F(\u) = \mathtt{TRUE} \}$
\end{description}
If one considers $\U$ a Uniform random vector on $\{0, 1\}^n$, then it is known \citep{rubinstein2011simulation} that:
\[
p_m = \proba{\U \in \mathcal{S} } = \dfrac{| \mathcal{S} |}{2^n} .
\]
Hence one can build an estimator of $| \mathcal{S} |$ by estimating the (extreme) probability $p_m$. In order to make use of the results of Section \ref{ss:Probability estimators}, one can consider the discrete random variable $X = S(\U)$ of the number of clauses satisfied by the assignment $\U$:
\begin{equation} \label{eq:definition X SAT}
X = S(\U) = \Sum[i = 1][m] S_i(\U)
\end{equation}
Therefore $X \in \llbracket 0, m \rrbracket$ and:
\[
\proba{\U \in \mathcal{S}} = \proba{F(\U) = \mathtt{TRUE}} = \proba{X = m} = \proba{X \geq m} .
\]

\subsubsection*{Conditional simulations} In order to perform the conditional simulations $\mu^X( \cdot \mid X \geq i)$ to simulate the random walks, we propose to use the Gibbs sampler as described in \citep{rubinstein2009gibbs} or \citep{cerou2011use}. Indeed, $\mu^X( \cdot \mid X \geq i) = \mu^U( \cdot \mid S(\U) \geq i )$. Starting from a sample $\U^*$ such that $S(\U^*) \geq i$, it resamples each coordinate in a deterministic (\emph{systematic} Gibbs sampler) or randomized order conditionally to the other ones to stay in the right domain. A sketch of a Gibbs sampler is given in Algorithm \ref{algo:gibbs sampler}.
\begin{algorithm}
\caption{Systematic Gibbs sampler}
\label{algo:gibbs sampler}
\begin{algorithmic}
\Require $\U = $
\State Draw $U_1 \sim \mu^U( \cdot \mid U_{2}, \cdots, U_n)$
\For{$i \gets 2, n-1$}
\State Draw $U_i \sim \mu^U( \cdot \mid U_1, \cdots,  U_{i-1}, U_{i+1}, \cdots, U_n)$
\EndFor
\State Draw $U_n \sim \mu^U( \cdot \mid U_1, \cdots, U_{n-1})$
\end{algorithmic}
\end{algorithm}
Practically speaking, to avoid local maxima and improve the convergence of the Markov chain, we do not start from the current $\U$ such that $X_i = S(\U)$. Instead, we pick at random a starting point $\U^*$ in a population already following the target distribution, \ie{} such that $S(\U^*) \geq X_i$. This population is built \emph{on-the-fly} with all the generated samples $\U$ such that $S(\U) \geq X_i$. Especially each \emph{fail} of a Geometric law will increase its size instead of replacing the previous vector as it is usual in a Multilevel Splitting method. We point out the fact that we focus on the real-valued random variable on the number of satisfied clauses and that multidimensional vectors are only seen as a way to generate it and to make conditional sampling on it. This makes us simulating the random walks by batches of size $k$ on the model of \citep{walter2015moving}: at a given iteration only the smallest \emph{states} are moved on. Reader is referred \citep{walter2015moving} for further details on parallel implementation.

Finally, we also generate both the strict and the non-strict random walks in the same run. This is to focus on the statistical properties of the number of iterations and of the estimators. Here some $\Delta_d = \proba{X > d}/\proba{X \geq d}$ are very small, so that the size of the population for the conditional sampling may become very \emph{small} if one only keeps those starting points strictly above the current threshold.

\subsubsection*{Numerical results}

\begin{figure}[!ht]
\centering
\subfloat[Strict random walk]{
\begin{knitrout}
\definecolor{shadecolor}{rgb}{0.969, 0.969, 0.969}\color{fgcolor}

{\centering \includegraphics[width=\maxwidth]{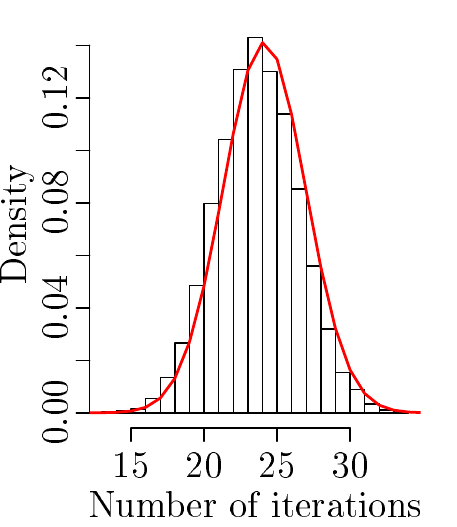} 

}

\end{knitrout}
\label{fig:niter strict sat}
}
\subfloat[Non-strict random walk]{
\begin{knitrout}
\definecolor{shadecolor}{rgb}{0.969, 0.969, 0.969}\color{fgcolor}

{\centering \includegraphics[width=\maxwidth]{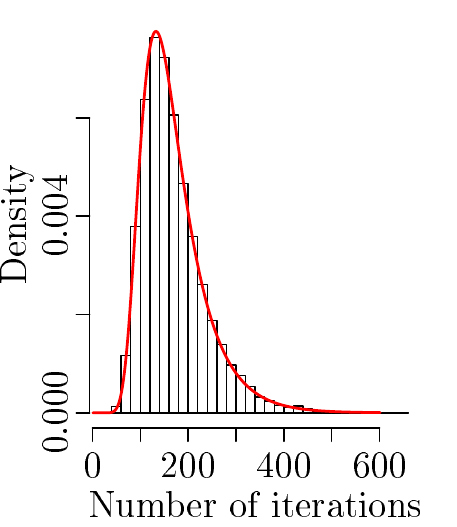} 

}

\end{knitrout}
\label{fig:niter nstrict sat}
}
\subfloat[Pure Poisson correction]{
\begin{knitrout}
\definecolor{shadecolor}{rgb}{0.969, 0.969, 0.969}\color{fgcolor}

{\centering \includegraphics[width=\maxwidth]{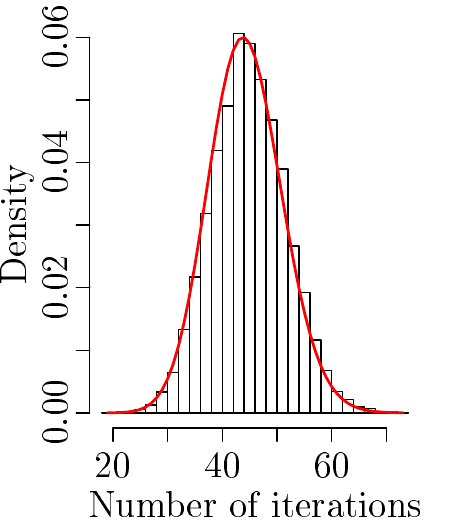} 

}

\end{knitrout}
\label{fig:niter poisson correction sat}
}
\caption{Histogram over $\ensuremath{10^{4}}$ realisations of the number of iterations for the strict random walk (\eqref{eq:loi M strict}), the non-strict one (\eqref{eq:loi M non strict}) and the pure Poisson \emph{correction} (Theorem \ref{th:generation M Y}). $X$ is an integer-valued random variable with up to $325$ jump points. The curves show the theoretical distributions with estimated Geometric parameters $\Delta_d$ with $N = 50000$.}
\label{fig:niter sat}
\end{figure}

We consider here the \emph{SAT} problem referred to as \texttt{uf75-01} on \texttt{satlib.org}, also used by \citet{botev2012efficient}, who provide a reference value $p = \ensuremath{5.977\times 10^{-20}}$ with a relative error of $0.03 \%$. It has $m = 325$ clauses in dimension $n = 75$. Hence $X$ is a discrete random variable with up to $325$ jump points. We first focus on the number of iterations of the random walks. To do so, we simulate $N = \ensuremath{10^{4}}$ random walks as well as the \emph{pure Poisson correction}. Figure \ref{fig:niter sat} plots the histograms of the random number of iterations for each case and theoretical curves with the $\Delta_d$ estimated using an other simulation with $N = 50000$.

These plots show a good consistency between theoretical formulae (\eqref{eq:loi M non strict} and \eqref{eq:loi M strict} and Theorem \ref{th:generation M Y}) and numerical results. Especially with a lot of jump points, the number of iterations are very different from Figure \ref{fig:niter strict sat} to Figure \ref{fig:niter nstrict sat} (almost hundred times bigger).

We now focus on the probability estimators. In this scope we also consider the \emph{Smoothed Splitting Method (SSM)} \citep{cerou2011use}, which uses a case-specific continuous auxiliary random variable. The aim of this benchmark is to assess the relevance of using such transformations instead of considering the original random variable with the MVUE we have proposed in Section \ref{ss:Probability estimators}. We refer the reader to \citep{cerou2011use} for further details on this transformation. The algorithm is then a usual Multilevel Splitting method with $p_0 = 0.2$. We set $N_\text{SSM}$ such that the total numbers of simulated samples for the non-strict random walk and for the SSM are of the same order of magnitude. The non-strict random walk generates on average $169.683$ samples while an AMS with $p_0 = 0.2$ typically generates $(1-p_0) N \log p / \log p_0$ samples. We set $N = 1000$ for the random walks, which gives $N_\text{SSM} \approx 7712$. Here we set $N_\text{SSM} = 8000$.

\begin{figure}[!ht]
\centering
\begin{knitrout}
\definecolor{shadecolor}{rgb}{0.969, 0.969, 0.969}\color{fgcolor}

{\centering \includegraphics[width=\maxwidth]{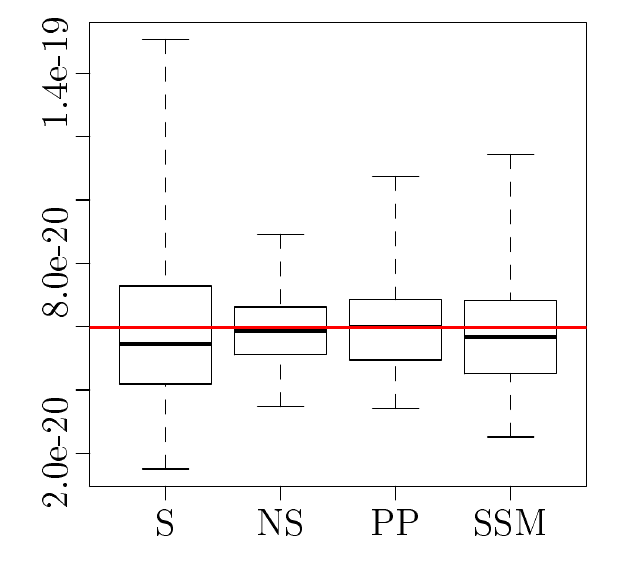} 

}

\end{knitrout}
\caption{Boxplots of the estimation of $p \approx \ensuremath{5.977\times 10^{-20}}$ over $100$ simulations with $N = 1000$, whiskers extending to the extreme values. (S): Strict random walk estimator (\eqref{eq:MVUE strict random walk}), (NS): Non-strict random walk estimator (\eqref{eq:MVUE non strict random walk}), (PP): Pure Poisson estimator (\eqref{eq:pure poisson estimator}), (SSM): Smoothed Splitting Method \citep{cerou2011use} with $N_\text{SSM} = 8000$.}
\label{fig:estimators sim sat}
\end{figure}

With a lot of discontinuity points, the variance differences between the three estimators is clearer, especially between the non-strict random walk estimator and the pure Poisson one. Also the strict inequality estimator has a much bigger variance. Indeed, for some discontinuity points $d \in D_x$, $\Delta_d \approx 10^{-2}$ while $N = 1000$, which gives coefficients of variations around $1/\sqrt{N p} \approx 32 \%$. This is not an issue in the non-strict case as the coefficient of variation of the MVUE of Lemma \ref{lem:MVUE Geom} typically scales like $\sqrt{(1-p)/N}$. On the other hand, over the $100$ simulations, we have an estimation of $\card (D_x) \approx 66$, which gives an upper bound for the squared coefficient of variation in the non-strict case (see \eqref{eq:bounds variance nstrict discret}): $0.068$, while the estimated squared coefficient of variation is $0.032$. Concerning the \emph{SSM} estimator, we have found a coefficient of variation of $0.33$ while the theoretical value should be $0.117$. As already noticed by \citep{cerou2011use} this is due to a non-perfect implementation of the Multilevel Splitting. This limitation is less visible while keeping the discrete random variable because we save all generated samples and so improve the approximation of the target distribution at each iteration.

Finally, these numerical results with a discrete random variable show a good consistency with the theoretical ones. Also it appears that it may not be relevant to transform the problem to consider a Multilevel Splitting method on a continuous random variable because this can make the conditional simulations harder to approximate. The pure Poisson \emph{correction} is in this context a good trade-off between accuracy and knowledge of the distribution. Furthermore, it is \emph{just} a by-product of the non-strict random walk and so the MVUE can also be computed in the same run. Concerning the strict inequality random walk, it may suffer from two limitations if the some $\Delta_d$ are very small (typically if $1/\Delta_d$ becomes of the order of the total population size $N$) because 
\begin{enumerate*}[label=\arabic*)]
\item the coefficient of variation of the MVUE of Lemma \ref{lem:MVUE bernoulli} is $\approx \sqrt{1/(N \Delta_d)}$, and
\item Markov chain drawing may be \emph{poor} because only few samples will be available in the right domain: at each iteration, only the samples strictly above the current threshold are kept, so on average only $N \Delta_d$ samples will be in the right domain. If Markov chain drawing is used to approximate conditional sampling, then the diversity of the population may decay strongly iterations after iterations.
\end{enumerate*}

\section{Conclusion}
The goal of this paper was to study the impact of using Multilevel Splitting methods on potentially discontinuous random variables, especially the parallel optimal (minimal variance) adaptive Multilevel Splitting method, also called the \emph{Last Particle Algorithm} which only replaces the smallest particle at a given iteration. This implementation had been shown to be a particular case of generating several point processes associated with a real-valued random variable, those point processes, also called increasing random walk here (see Section \ref{ss:The increasing random walk}) being related to a Poisson process with parameter 1. To handle potential discontinuities, we had to distinguish between strict and non-strict inequality for conditional sampling. Following \citet{simonnet2014combinatorial} who gave the law of the number of iterations of such algorithms with non-strict inequality, we have been able to extend this result to the strict inequality random walk. Furthermore, we have given the Minimal Variance Unbiased Estimator of the probability of exceeding a threshold in both cases. These estimators eventually become the same when the random variable of interest is actually continuous. On the other hand their distributions are not easy to characterise. In this scope, we have proposed a third estimator based on the non-strict inequality implementation. It has always the same statistical properties, $X$ being continuous or not, precisely the properties for continuous random variables recalled in Section \ref{ss:The increasing random walk}. These properties are well known and allow for building confidence intervals on the estimator. All these theoretical results have been challenged on two examples and proved good agreement with the results.

Practically speaking, the strict inequality implementation may be less efficient if conditional sampling has to be done with Markov chain drawing and can even return $0$ if some jump points $d \in D_x$ are such that $\Delta_d^{-1} = \proba{X \geq d}/\proba{X > d}$ is of the order of the population size:  This is due to the estimation of the $\Delta_d$ with crude Monte Carlo. The non-strict implementation prevents from such issues because it generates Geometric random variables. On the other hand there is no literal expression of the variance of its estimator. At the cost of generating independent Uniform random variables, the non-strict implementation can also output the \emph{pure Poisson} estimator, it is the estimator with known distribution which does not depend on the discontinuities of $X$. If one only want to focus on the output without worrying about discontinuities, this latter appears as good trade-off.

Finally, considering the global cost of an estimator against its variance, some optimisation may be done to start some random walks only from a given point and/or stop them before the targeted value because all the jumps are not of equal size. This has not been studied here and is let for further research.

\begin{acknowledgement}
The author would like to thank his advisors Josselin Garnier (University Paris Diderot) and Gilles Defaux (Commissariat \'a l'\'Energie Atomique et aux \'Energies Alternatives) for their advices and suggestions.
\end{acknowledgement}

\appenproof
\Closesolutionfile{ann}
\begin{demprop}{1.6}
The distribution of $M_x$ has already been proved by \citep{simonnet2014combinatorial} assuming that $\card D_x < \infty$. We extend this result to the possible infinite countable number of discontinuity.

Let $S_n = \{ x \in \R \mid 0 < \proba{X = x} < 1/n \}$ be the set of the jump points of $F_X$ with jump amplitudes smaller than $1/n$ and $X^{(n)} = X \one_{X \notin S_n}$. $X^{(n)}$ has a finite number of jump points and accumulates the (possibly infinite) number of jump points $d \in D \mid \proba{X = d} < 1/n$ at $0$. Since it has a finite number of discontinuities, the law of its associated counting random variable is then known.

Furthermore it verifies:
\[
\forall x \in \R,\; \proba{X^{(n)} \leq x} = \proba{X \leq x} + \one_{x \geq 0} \Sum[\substack{d \in S_n \\ d > x}][] \proba{X = d} - \one_{x<0} \Sum[\substack{d \in S_n \\ d \leq x}][] \proba{X = d}.
\]
Hence $X^{(n)} \xrightarrow[n \to \infty]{\mathcal{L}} X$, which implies $M_x^{(n)} \xrightarrow[n \to \infty]{\mathcal{L}} M_x$ with $M_x^{(n)}$ the counting random variable at state $x$ associated with an increasing random walk on $X^{(n)}$.

Moreover, one has: $\forall x \in \R,\; \forall n \geq 1,\; M_x^{(n)} \sim \mathcal{P} \l -\log \dfrac{\proba{X^{(n)} > x}}{\prod \limits_{d \in D_x \setminus S_n} \Delta_d} \r + \Sum[d \in D_x \setminus S_n][] \mathcal{G}\l \Delta_d \r$. Finally, this gives:
\[
M_x \sim \mathcal{P} \l -\log \dfrac{\proba{X > x}}{\prod \limits_{d \in D_x} \Delta_d} \r + \Sum[d \in D_x][] \mathcal{G}\l \Delta_d \r .
\]
We also show that the first and second order moments of $M_x$ remain finite even when $\operatorname{card}(D_x) = \infty$. One has:
\begin{align*}
0 \leq \Sum[d \in D_x][] \l \dfrac{1}{\Delta_d} - 1 \r = \Sum[d \in D_x][] \dfrac{\proba{X = d}}{\proba{X > d}}
\leq \dfrac{1}{\proba{X > x}} \Sum[d \in D_x][] \proba{X = d} \leq \dfrac{1-p_x}{p_x},
\end{align*}
and:
\[
1 \geq \prod \limits_{d \in D_x} \Delta_d \geq \prod \limits_{d \in D_x} e^{-\l \frac{1}{\Delta_d} - 1 \r} \geq e^{-\Sum[d \in D_x][] \l \frac{1}{\Delta_D} - 1 \r} \geq e^{-\frac{1-p_x}{p_x}} > 0.
\]
All together, theses inequalities give the result:
\begin{align*}
\E{M_x} &= -\log \dfrac{p_x}{\prod \limits_{d \in D_x} \Delta} + \Sum[d \in D_x] \l \dfrac{1}{\Delta_d} - 1 \r \leq -\log p_x + \dfrac{1 - p_x}{p_x} \\
\var{M_x} &= -\log \dfrac{p_x}{\prod \limits_{d \in D_x} \Delta} + \Sum[d \in D_x] \dfrac{1}{\Delta_d} \l \dfrac{1}{\Delta_d} - 1 \r \leq -\log p_x + \dfrac{1 - p_x}{p_x^2} .\\
\end{align*}
\end{demprop}
\begin{demprop}{1.7}
Using the renewal property of the Poisson Process the number of events in the \emph{continuous} part will be the same as the one in the non-strict case. Indeed the only difference with the non-strict random walk comes from the \emph{behaviour} of the random walk when $X_n \in D_x$. In this latter case, while the non-strict inequality repeats the trial until success, the strict inequality do it only once. Hence the Geometric law is replaced by a Bernoulli one.
\end{demprop}
\begin{demlem}{1.8}
One is going to use Lehmann-Scheffé theorem with the statistic $T = \Sum G_i$. As the sum of $N$ independent Geometric random variables with parameter $p$, $T$ follows a Negative Binomial law: $\forall t \in \N, \; \proba{T = t} = \binom{N+t-1}{t} p^N (1-p)^t$. $T$ is sufficient:
\[
\mathcal{L}(g_1, \cdots, g_N, p) = \prod \limits_{i = 1}^{N} \proba{G_i = g_i} = (1-p)^{\Sum g_i} p^N = (1-p)^t p^N .
\]
$T$ is also complete: let $\phi$ be a measurable function, one has:
\begin{align*}
\forall p \in (0,1),\; \E{\phi(T)} = 0 &\Rightarrow \forall p \in (0,1),\; p^N \Sum[t = 0][\infty] \binom{t + N -1}{t} \phi(t) (1-p)^t = 0 \\
&\Rightarrow \forall \theta \in (0,1),\; \Sum[t = 0][\infty] \alpha_t \theta^t
\end{align*}
with $\alpha_t = \binom{t + N - 1}{t} \phi(t)$ and $\theta = 1-p$. Furthermore $p=1$ \ie{} $\theta = 0$ gives $\phi(0) = 0$ and $\theta = 1$, \ie{} $p = 0$ gives $\proba{T < \infty} = 0$. Hence the power series $\theta \mapsto \sum \alpha_t \theta^t$ is identically null on its radius of convergence $[0, 1)$ and so $\forall t \in \N, \; \alpha_t = 0$, which means $\forall t \in \N,\; \phi(t) = 0$ and $T$ is complete.

We now consider the estimator $R = \dfrac{1}{N} \Sum \one_{G_i=0}$. $R$ is unbiased because $\E{\one_{G_i=0}} = \proba{G_1 = 0} = p$. Then Lehmann-Scheffé theorem states that $\E{R \mid T}$ is the MVUE of $p$. This gives:
\begin{align*}
\E{R \mid T = t} &= \dfrac{1}{N} \Sum \E{ \one_{G_i=0} \mid \Sum G_i = t} = \proba{G_1 = 0 \mid \Sum G_i = t} \\
&= \dfrac{\proba{G_1 = 0, \Sum[i=2][N] G_i = t}}{\proba{\Sum G_i = t}} = \proba{G_1 = 0} \dfrac{\binom{t + N - 2}{t} p^{N-1} (1-p)^t}{\binom{t + N - 1}{t} p^{N} (1-p)^t} \\
\E{R \mid T = t} &= \dfrac{N-1}{N-1 + t}
\end{align*}
Hence, $\p = \E{R \mid T} = (N-1)/(N-1 + T)$ is the MVUE of $p$.
\end{demlem}
\begin{demprop}{1.11}
On the one hand, for all $a<b$ such that $X$ is continuous on $(a, b)$, $\proba{X > b \mid X > a}$ can be estimated by $(1-1/N)^{\# \{ X_n \in (a,b) \}}$ with $\# \{ X_n \in (a,b) \}$ the number of events of the sum process in $(a,b)$. Moreover, since $b \mapsto \proba{X \geq b}$ is left-continuous, it is also an MVUE of $\proba{X \geq b \mid X > a}$ and this relation remains true if $b \to a$ since $\operatorname{card}(\varnothing) = 0$. Using the fact that the RLE of $(X_1, \cdots, X_{M_x})$, $\mathbf{r} = (1, ..., 1)$ with probability $1$ when $X$ is continuous, $(1-1/N)^{\# \{ X_n \in (a,b) \}} = \prod (N-1)/(N-1 + r_i)$ with probability $1$.

On the other hand $\forall d \in D_x, \; \Delta_d = \proba{X > d}/\proba{X \geq d}$ can be estimated with the MVUE defined in Lemma \ref{lem:MVUE Geom}. The first state of each chain non smaller than $d$ can be considered as an \iid sample of $X \mid X \geq d$. The number of times $d$ is found in $(X_1, \cdots, X_{M_x})$ is then the sum of $N$ realisations of a Geometric random variable with parameter $\Delta_d$. Hence, $\Delta_d$ is estimated with $(N-1)/(N-1+r_d)$ with $r_d$ the number of times $d$ is found in $(X_1, \cdots, X_{M_x})$.

Since $D$ is countable, we consider $\R \setminus D = \bigcup_i \, I_i$ with $(I_i)_i$ a sequence of disjoint open intervals. Note that some subsequence of $I_n$ may converge toward the empty set if $D$ is infinite countable with some accumulation points. However $X$ is continuous on $\R \setminus D$. Using the renewal property of the Poisson process, one can consider that $\prod (N-1)/(N-1 + r_i)$ is a product of independent MVUE estimators. Especially, denoting by $M_\text{pois}$ the number of $1$ in $\mathbf{r}$, $(1-1/N)^{M_\text{pois}}$ is a MVUE of $p_\text{pois} := p_x / \prod \Delta_d$.

We now explicit the calculation for the bounds on the variance. One has:
\begin{align*}
\E{\p_x^2} = p_\text{pois}^2 p_\text{pois}^{-1/N} \prod \limits_{d \in D_x} \E{ \l \dfrac{N-1}{N-1 + T_d} \r^2}
\end{align*}
with $T_d \sim NegBin(N, \Delta_d)$. For a given $d \in D_x$, one has:
\begin{align*}
\E{\l \dfrac{N-1}{N-1 + T_d} \r^2} &= \Sum[t = 0][\infty] \binom{t + N - 1}{t} (1-\Delta_d)^t \Delta_d^N \l \dfrac{N-1}{N-1+t} \r^2 \\
&= \Sum[t = 0][\infty] \binom{t + N - 2}{t} (1-\Delta_d)^t \Delta_d^N \dfrac{N-1}{N-1+t} \\
&= \Delta_d \Sum[t = 0][\infty] \binom{t + N - 2}{t} (1-\Delta_d)^t \Delta_d^{N-1} \dfrac{N-2}{N-2+t} \dfrac{N-1}{N-2} \dfrac{N-2+t}{N-1+t}.
\end{align*}
Furthermore, $\forall t \geq 0,\; \dfrac{N-1}{N-2} \dfrac{N-2+t}{N-1+t} \in [1, (N-1)/(N-2)]$, which gives:
\[
\Delta_d^2 \leq \E{\l \dfrac{N-1}{N-1 + T_d} \r^2} \leq \Delta_d^2 \dfrac{N-1}{N-2}.
\]
Eventually the variance writes:
\begin{align*}
p_x^2 \l p_\text{pois}^{-1/N} - 1 \r \leq \var{\p_x} &= \E{\p_x^2} - p_x^2
%\leq p_\text{pois}^{2-1/N} \prod \limits_{d \in D_x} \Delta_d^2 \dfrac{N-1}{N-2} - p_x^2 = p_x^2 \l p_x^{-1/N} \prod \limits_{d \in D_x} \Delta_d^{1/N} \dfrac{N-1}{N-2} - 1 \r
\leq p_x^2 \l p_\text{pois}^{-1/N} \prod \limits_{d \in D_x} \dfrac{N-1}{N-2} - 1 \r .
\end{align*}
\end{demprop}
\begin{demprop}{1.12}
The same reasoning	as for the proof of Proposition \ref{propo:MVUE non strict random walk} applies where the MVUE of a Geometric law is replaced by the one of a Bernoulli distribution. For the variance one has:
\begin{align*}
\E{\p_x^2} &= p_\text{pois}^2 \; p_\text{pois}^{-1/N} \prod \limits_{d \in D_x} \dfrac{\Delta_d^2}{N} \l N-1 + \dfrac{1}{\Delta_d} \r \\
\E{\p_x^2} &= p_x^2 \; p_x^{-1/N} \prod \limits_{d \in D_x} \dfrac{\Delta_d (N-1) + 1}{N \Delta_d^{1-1/N}}
\end{align*}
which gives the result.
\end{demprop}
\begin{demtheo}{1.13}
The difference between the generation of the non-strict random walk and Algorithm \ref{algo:pseudo code Y} stands in the addition of the \emph{while} loop from line \ref{line:debut while test X in D} to line \ref{line:fin while test X in D}. This loop is entered when a Geometric scheme is started: two consecutive events being equal is a non-zero probability event only when $X_n \in D$. In this context, the condition $\{U_{n+1} > U_n \}$ lets generate the counting random variable of the continuous random variable associated with the Geometric scheme as explained in Section \ref{ss:Probability estimators}. Therefore it follows a Poisson distribution with parameter $- \log \Delta_{X_n}$. The renewal properties of the Poisson Process lets conclude the proof.
\end{demtheo}

\bibliographystyle{apalike}
\bibliography{biblio}

\end{document}